\documentclass{webofc}

\usepackage[varg]{txfonts}   
\usepackage{hyperref}
\usepackage{url}
\hypersetup{colorlinks=true,citecolor=blue,urlcolor=blue,linkcolor=blue}
\begin{document}
\title{New collider implications
on a strongly first order EWPT\thanks{Presented at the International Workshop on Future Linear Colliders LCWS2024\qquad\qquad\qquad OU-HET-1245}}
%
%

\author{\firstname{Ricardo} \lastname{R. Florentino}\inst{1}\fnsep\thanks{Speaker \email{rflorentino@hetmail.phys.sci.osaka-u.ac.jp}} \and
        \firstname{Shinya} \lastname{Kanemura}\inst{1}\fnsep\thanks{\email{kanemu@het.phys.sci.osaka-u.ac.jp}} \and
        \firstname{Masanori} \lastname{Tanaka}\inst{2}\fnsep\thanks{\email{tanaka@pku.edu.cn}}
}

\institute{Department of Physics, Osaka University, Toyonaka, Osaka 560-0043, Japan 
\and
           Center for High Energy Physics, Peking University, Beijing 100871, China
          }

\abstract{In order to understand the early history of the universe, and to test baryogenesis models, determining the nature of the electroweak phase transition is imperative. The order and strength of this transition is strongly correlated to relatively large deviations in the $hhh$ coupling. In models where a considerable part of the $hhh$ coupling deviation is caused by charged particle loops, the $h\gamma\gamma$ coupling is also expected to deviate considerably. In this talk, by using a model-independent approach, I explain how to obtain conditions that are sufficient for a strongly first order phase transition. After the $h\gamma\gamma$ coupling is determined with precision at the HL-LHC, these conditions can be tested at Future Linear Colliders by measurements of the $hhh$ coupling, to conclusively determine the nature of the electroweak phase transition and the viability of electroweak baryogenesis on models with new charged scalars.}
\maketitle

\section{Introduction}

The Standard Model (SM) is extremely successful in explaining most currently observed phenomena. Nonetheless, one of its main problems is the absence for an explanation of the observed Baryon Asymmetry of the Universe (BAU)\cite{Planck:2018vyg}. One of the most promising methods of baryon generation is known as Electroweak Baryogenesis (EWBG)\cite{Kuzmin:1985mm}. This kind of baryogenesis is possible when the Electroweak Phase Transition (EWPT), which occurs in the early universe at the electroweak scale, is a Strongly First Order Phase Transition (SFOPT)\cite{Sakharov:1967dj}. In the Higgs sector of the SM, the EWPT is not a SFOPT\cite{Kajantie:1996mn,DOnofrio:2014rug,DOnofrio:2015gop}. Therefore, Beyond the SM (BSM) models with extended Higgs sectors are necessary for a successful EWBG.

The occurrence of a SFOPT in the context of EWBG has various experimental consequences. In this work these consequences are studied, both at colliders and in cosmology, using a model independent framework known as the nearly-aligned Higgs Effective Field Theory (naHEFT). Furthermore, it is discussed how the measurement of the $h\to\gamma\gamma$ decay at future experiments can be used to constrain the allowed parameter space for a SFOPT in given models, which will be available considerably sooner than triple Higgs coupling measurements. Once the triple Higgs coupling is measured with precision in e.g. the latter stages of the ILC, the $h\gamma\gamma$ will be a crucial complement in pinpointing the models capable of generating a SFOPT.

This work is organised as follows. In section 2 the motivation and formalism of the naHEFT is presented. In section 3 the cosmological constraints on a SFOPT, as well as the cosmological observables considered, are discussed. In section 4 the main results are shown as figures constraining the parameter space of various benchmark models. The situation of future collider physics is further explored in section 5, with a focus on the International Linear Collider (ILC) program. Finally, conclusions are given in section 6.

\section{The nearly-aligned Higgs EFT}

The two main description for BSM physics within an Effective Field Theory (EFT) approach are the Standard Model EFT (SMEFT)\cite{Buchmuller:1985jz,Hagiwara:1993ck,Grzadkowski:2010es} and the Higgs EFT (HEFT)\cite{Feruglio:1992wf,Alonso:2012px,Brivio:2013pma,Buchalla:2013rka,Buchalla:2017jlu,Falkowski:2019tft,Cohen:2020xca,Sun:2022ssa,Sun:2022snw}. The former is based on canonical dimension counting, and thus reliable in the decoupling limit. The latter is based on chiral dimension counting, and thus of ideal use in the non-decoupling scenario\cite{Banta:2021dek, Banta:2022rwg, Buchalla:2023hqk}. In the context of EWPT, non-decoupling effects are crucial to produce a SFOPT. HEFT is then the correct framework to study such phenomenology.

In the most general HEFT, infinite operators are generated, even at leading order, so that HEFT by itself has very little prediction power. The naHEFT is based on the HEFT, with the assumption that all deviations to Higgs boson couplings are generated at loop level\cite{Kanemura:2022txx,Kanemura:2021fvp}. This assumption is motivated by the collider measurements being incredibly close to SM predictions. This allows the Lagrangian to take the form of one-loop effects of heavy particles integrated out. The Higgs potential will then assume a Coleman-Weinberg like form, and the resulting Lagrangian will be of the form:

\begin{equation}
    \mathcal{L}_{naHEFT}=\mathcal{L}_{SM}+\xi(\mathcal{L}_{S}+\mathcal{L}_{V})\qquad\qquad\left(\xi=\frac{1}{(4\pi)^2}\right)
\end{equation}

\noindent where $\mathcal{L}_{S}$ is the scalar sector:

\begin{equation}
    \mathcal{L}_{S}=-\frac{\kappa_0}{4}[\mathcal{M}^2(h)]^2\log\frac{\mathcal{M}^2(h)}{\mu^2}+\frac{v^2}{2}\mathcal{F}(h)Tr[D_\mu U^\dagger D^\mu U]+\frac{1}{2}\mathcal{K}(h)(\partial_\mu h)(\partial^\mu h)
\end{equation}

\noindent and $\mathcal{L}_{V}$ is the scalar-vector potential important to describe the loop generated $h\to\gamma\gamma$ decay:

\begin{equation}
    \mathcal{L}_{V}=g^2\mathcal{F}_W(h)\text{Tr}[\mathbf{W}_{\mu\nu}\mathbf{W}^{\mu\nu}]+g'^2\mathcal{F}_B(h)\text{Tr}[\mathbf{B}_{\mu\nu}\mathbf{B}^{\mu\nu}]-gg'\mathcal{F}_{BW}(h)\text{Tr}[U\mathbf{B}_{\mu\nu}U^{\dagger}\mathbf{W}^{\mu\nu}]
\end{equation}

The non-decouplingness of a particle is given by the fraction of their mass that is generated from the SM Higgs boson, $r=v^2\lambda_p/\Lambda$, where $\Lambda$ is the particles mass and $\lambda_p$ it's coupling with the SM Higgs boson. In this work, a further simplification is made, that the mass scale and the non-decouplingness of the BSM particles are approximately degenerate, such that the new physics can be described by three main parameters: the mass scale $\Lambda$, the non-decouplingness $r$, and the degrees of freedom $k_0$. The last of which can be further subdivided in the degrees of freedom for scalars of each charge, for the purpose of the $h\to\gamma\gamma$ calculation. The polynomials of the Higgs boson present in the Lagrangian can in this case be described by:

\begin{align}
    \mathcal{F}(h)=&\mathcal{F}_{BW}(h)=0\\
    \mathcal{M}^2(h)=&M^2+\lambda_g(v+h)^2\\
    \mathcal{K}(h)=&\kappa_0\frac{\Lambda^2}{3v^2}r\left[1-(1-r)\frac{\Lambda^2}{\mathcal{M}^2(h)}\right]\\
    \mathcal{F}_W(h)=&\mathcal{F}_B(h)=\frac{b}{2}\ln\left[1-r+r\left(1+\frac{h}{v}\right)^2\right],\quad b=\frac{n_++4n_{++}}{3}
\end{align}

Implementing these function in the Lagrangian, it is possible to deduce the following coupling scaling factors:

\begin{align}
    &\kappa_{V} = \kappa_{f} =  1 - \kappa_{0} \frac{\xi}{6} \frac{\Lambda^2}{v^2} r^2, \\
    &\kappa_{3} = 1 + \kappa_{0} \frac{4 \xi}{3} \frac{\Lambda^4}{v^2 m_{h}^2} \left[ r^3 - \frac{m_{h}^2}{8 \Lambda^2} r^2 (3-2r)\right], \\
    &\kappa_{\gamma \gamma}^2 \simeq \left| \kappa_{V} -  \frac{br}{F_{\rm SM}}  \right|^2, \\
    &\kappa_{Z \gamma}^2 \simeq \left| \kappa_{V} -  \frac{br}{G_{\rm SM}} \left( J_{3}^{\rm new} - s_{W}^2 \right)  \right|^2, 
\end{align}

\noindent these represent deviations from the standard model in the Higgs coupling with gauge/fermion pairs, in the tripple Higgs coupling, and in the Higgs decay to $\gamma\gamma$ and $Z\gamma$ respectively. A point to note is that most deviations are proportional to powers in $\kappa_0$, $r$, and $\Lambda$, while the deviations in the $\kappa_{\gamma\gamma}$ is proportional to $b=(n_++4n_{++})/3$, leading to a different, model dependent behaviour.

\section{Cosmology of a Strongly First Order EWPT}

In order to have a SFOPT and successfully generate the BAU, some cosmological requirements are necessary to employ. The first one says that the sphaleron process must decouple almost instantly after the phase transition. This is often refered to as the sphaleron decoupling condition\cite{Kuzmin:1985mm}, and can be approximated by:

\begin{equation}
    \frac{v_n}{T_n}>1
\end{equation}

\noindent where $v_n$ is the Higgs vacuum at the nucleation temperature $T_n$ of the phase transition. This condition is strongly correlated with the strength of the phase transition, and thus, with the deviations in the triple Higgs coupling\cite{Grojean:2004xa, Kanemura:2004ch, Kakizaki:2015wua, Hashino:2016rvx}.

The second condition says the transition rate of the phase transition must be high enough compared to the expansion of the universe for the transition to complete by today. This is referred to as the completion condition\cite{Turner:1992tz}, and can be approximated by:

\begin{equation}
    \frac{\Gamma}{H^4}>1
\end{equation}

\noindent where $\Gamma$ is the transition rate and $H$ the Hubble parameter.

Being of cosmological nature, a SFOPT also has cosmological observable that can be observed at various experiments. One of them is possible gravitational waves (GW) generated during the phase transition\cite{Grojean:2006bp, Kakizaki:2015wua, Hashino:2016rvx,Hashino:2018wee}. In general, cosmological phase transitions have three sources of gravitational waves: bubble collision, sound waves, and plasma turbulence:

\begin{equation}
    h^2 \Omega_{\rm GW}(f) \simeq h^2 \Omega_{\varphi}(f) + h^2 \Omega_{\rm sw}(f) +  h^2 \Omega_{\rm turb}(f), 
\end{equation}

\noindent The fitting functions used for the GW spectrum $ \Omega_{\rm GW}(f)$ are shown in Ref.~\cite{Caprini:2015zlo}. The GW are considered detectable if their signal to noise ratio is larger than ten\cite{Cline:2021iff}, $SNR>10$, where

\begin{equation}
    {\rm SNR} = \sqrt{ \mathcal{T} \int^{f_{\rm max}}_{f_{\rm min}} df \left[ \frac{h^2 \Omega_{\rm GW}(f)}{h^2 \Omega_{\rm sens}(f)} \right]^2 }, 
\end{equation}

\noindent and where $\mathcal{T} = 1.26 \times 10^{8}\,{\rm s}$ is used for detectability at both LISA\cite{LISA:2017pwj} and DECIGO\cite{Kawamura:2011zz} interferometers.

Finally, the production and detectability of Primordial Black Holes is considered. These are generated in the scenario where a delay of the phase transition happens across a Hubble volume, and a large density contrast is present, leading to the creation of PBH\cite{Hashino:2021qoq,Liu:2021svg}. The same analysis as in \cite{Hashino:2022tcs} is performed. The PBH are considered detectable if the fraction of PBH is larger than $10^{-4}$ such that it be detect in experiments like Subaru HSC~\cite{Niikura:2017zjd}, OGLE~\cite{Niikura:2019kqi}, PRIME~\cite{Kondo_2023} and Roman telescope~\cite{fardeen2023astrometric}. 

\section{Parameter Space Examination}

Here we summarise some of the main results presented previously in~\cite{Florentino:2024kkf}.

\begin{figure}[ht]
\centering
\includegraphics[width=\textwidth]{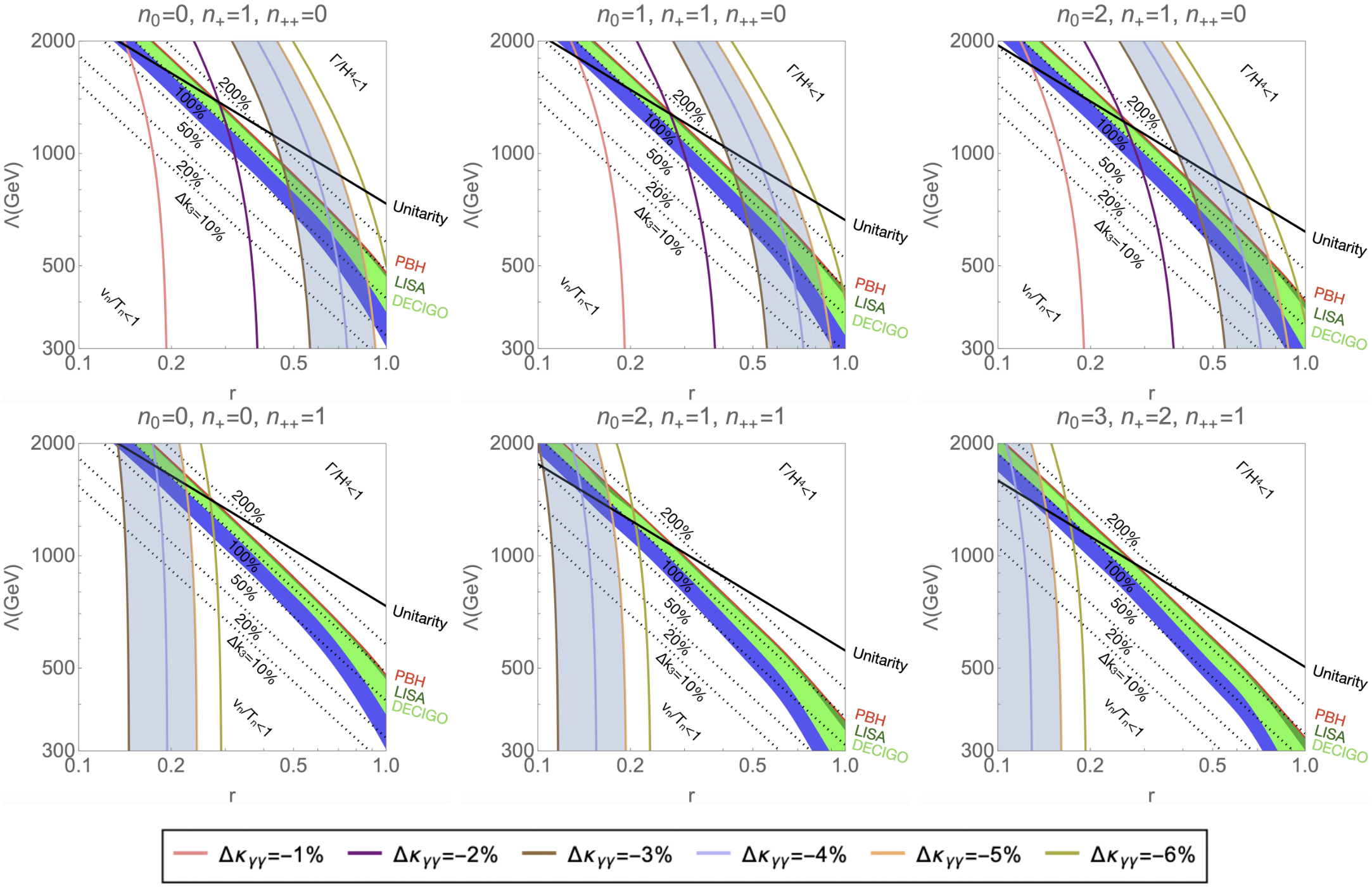}
\caption{Parameter region in the $(\Lambda,r)$ space for various possibilities of scalar degrees of freedom corresponding to some benchmark models. The white region cannot satisfy one of the conditions for a SFOPT. The light green region produces gravitational waves that can be tested at DECIGO. The gravitational waves in the dark green region can also be detected at LISA. The red region also generates PBH detectable in the considered experiments. The blue region satisfies the SFOPT, but can only be probed at colliders. The dashed lines represent contours in $\kappa_3$. The remaining coloured lines represent contours in the $\kappa_{\gamma\gamma}$. The transparent blue region represents an illustrative scenario where $\Delta\kappa_{\gamma\gamma}=(-4\pm1)\%$ is measured at future colliders.}
\label{fig-1}
\end{figure}

In figure~\ref{fig-1} the parameter region studied is shown. Each graph corresponds to different number of scalar degrees of freedom. These correspond to various know BSM models with extend scalar fields like two Higgs doublet models~\cite{Arhrib:2003vip,Aoki:2009ha,Posch:2010hx,Arhrib:2012ia,Chiang:2012qz,Fontes:2014xva,Kanemura:2014dja,Kanemura:2014bqa,Arhrib:2015hoa,Kanemura:2015mxa,Hashino:2015nxa,Kanemura:2016sos,Senaha:2018xek,Braathen:2020vwo,Florentino:2021ybj,Degrassi:2023eii,Aiko:2023nqj,Aiko:2023xui}, Higgs triplet models~\cite{Akeroyd:2010je,Arhrib:2011vc,Aoki:2012yt,Kanemura:2012rs,Aoki:2012jj,Arbabifar:2012bd,Chiang:2012qz}, the SM with singlet scalar fields~\cite{Shifman:1979eb, Chiang:2012qz,Katz:2014bha,Kakizaki:2015wua,Kanemura:2015fra,Kanemura:2016lkz,Braathen:2019pxr,Braathen:2019zoh,Aiko:2023xui}, and the Georgi-Machacek model~\cite{Georgi:1985nv, Chiang:2017vvo,Chiang:2018xpl}. The BSM mass scale $\Lambda$ and non-decouplingness $r$ are scanned for compliance with the SFOPT conditions and detection at GW interferometers and PBH observations. The white area below the coloured region violates the sphaleron decoupling condition, while the white area above the coloured region violates the phase transition completion condition. The blue, light and dark green, and red regions can thus realise a SFOPT. The blue area can only be probed at colliders, while the light green area and above can generate GW detectable at DECIGO, the dark green area and above can be probed at LISA, and the red area can also produce detectable PBH. The dashed lines represent contrours of the triple Higgs coupling scaling factor $\kappa_3$ while the other coloured lines represent contours in the $h\to\gamma\gamma$ decay scaling factor $\kappa_{\gamma\gamma}$. Finally, the transparent blue area represents the area where $\Delta\kappa_{\gamma\gamma}=(-4\pm1)\%$.

From figure~\ref{fig-1} it can be seen that while LISA and PBH can only probe a small section of the allowed parameter space, DECIGO is able to probe a larger area. On the other hand, the triple Higgs coupling is shown to cover all the area, and as expected, a strong determination of $\kappa_3$ is essential for the determination of the nature of the EWPT. Further details on the capacity of future $\kappa_3$ measurements to determine the nature of the SFOPT are presented in the next section. The $\kappa_{\gamma\gamma}$ contours, by being unaligned with the other measurements, reveal the $h\to\gamma\gamma$ measurements can be used as a strong complement to the former to test the nature of the EWPT. As an illustrative example, the possibility of future experiments, like the $HL-LHC$, measuring $\Delta\kappa_{\gamma\gamma}=(-4\pm1)\%$ is considered. In that case, only the transparent blue regions in figure~\ref{fig-1} are allowed. Thanks to the verticality of these lines, the parameter for a SFOPT reduces immensely, in a very model dependent way. The $\kappa_{\gamma\gamma}$ can then put strong model-dependent bounds on the parameters. In the $(n_0,n_+,n_{++})=(0,1,0)$ case, corresponding to the real singlet extension of the SM, these bounds can be determined as $\Lambda < 946\,{\rm GeV}$ and $r>0.457$, with the triple Higgs coupling being constrained to $137\,\%>\Delta \kappa_{3}> 21 \%$.

\begin{table}
\centering
\caption{In the ilustrative scenario where $\Delta\kappa_{\gamma\gamma}=(-4\pm1)\%$ is measured, bounds on the $\kappa_3$ are presented for each of the models in figure~\ref{fig-1}. The bounds in the second column represent intervals for which the EWPT can be strongly first order, meaning measurements outside this range conclusively exclude a SFOPT. The bounds in the third column represent the range for witch the EWPT is strongly first order, meaning that measurements within this range conclusively confirm a SFOPT.}
\label{tab-1}       
\begin{tabular}{|c|c|c|c|}
\hline 
~$(n_{0},\, n_{+}, \, n_{++})$~  &  ~Required by SFOPT~ & ~Conservative bound~ & Example of SM extension
\\ \hline 
$(0, \, 1, \, 0)$ & $137\,\%>\Delta \kappa_{3}> 21 \%$ & ~$114\,\%>\Delta \kappa_{3}> 50\,\%$~ & ~A singly charged scalar~
\\ \hline
$(1, \, 1, \, 0)$ & ~$143\,\%>\Delta \kappa_{3}> 19\,\%$~ & $115\,\%>\Delta \kappa_{3}> 47\,\%$ & ~A real triplet scalar~
\\ \hline
$(2, \, 1, \, 0)$ & ~$135\,\%>\Delta \kappa_{3}> 18\,\%$~ & $114\,\%>\Delta \kappa_{3}> 44\,\%$ &~A doublet scalar~
\\ \hline 
$(0, \, 0, \, 1)$ & ~$153\,\%>\Delta \kappa_{3}> 62\,\%$ & $148\,\%>\Delta \kappa_{3}> 65\,\%$ & ~A doubly charged scalar~
\\ \hline
$(2, \, 1, \, 1)$ & ~$160\,\%>\Delta \kappa_{3}> 65\,\%$~ & $150\,\%>\Delta \kappa_{3}> 75\,\%$ & ~A complex triplet scalar~
\\ \hline
$(3, \, 2, \, 1)$ & ~$136\,\%>\Delta \kappa_{3}> 59\,\%$~ & $153\,\%>\Delta \kappa_{3}> 63\,\%$ &~Gerogi-Machacek model~
\\ \hline
\end{tabular}
\end{table}

Focusing on the triple Higgs coupling, table~\ref{tab-1} explores the type of bounds one can obtain. Still in the illustrative example of $\Delta\kappa_{\gamma\gamma}=(-4\pm1)\%$ being measured in future experiments, the same benchmark models in figure~\ref{fig-1} are explored. Two different bounds on $\kappa_3$ are shown. In the second column, a bound in which both the sphaleron decoupling condition and the completion condition are satisfied is shown. Therefore, inside this interval there can be SFOPT, but it is not guaranteed to happen. A measurement of $\kappa_3$ outside this interval can then conclusively deny the capacity of the respective model to explain a SFOPT. The third column, on the other hand, shows the interval for which both the sphaperon decoupling and the completion conditions are guaranteed to be satisfied. This means that a measure inside this interval can conclusively confirm the occurrence of a SFOPT in the respective model. This motivates how $h\to\gamma\gamma$ measurements will give us precious insight on the nature of the EWPT in various models before the triple Higgs coupling is measured at future colliders.

\section{Discussion on Future Colliders}

Here further motivation of the importance of $h\to\gamma\gamma$ measurements in determining the nature of the EWPT is provided, by exploring the expected measurements of the triple Higgs coupling at future colliders and consider their implication on the discussion presented previously.

\begin{table}
    \centering
    \caption{Expected precision at one sigma of the triple Higgs coupling measurements at the HL-LHC and various stages of the ILC.}
    \label{tab-2}
    \begin{tabular}{c|c|c|c|c}
        Collider & HL-LHC & ILC$_{250}$ & ILC$_{500}$ & ILC$_{1000}$ \\\hline
        $\Delta\kappa_3(1\sigma)$ & 50\% & 49\% & 22\% & 10\%
    \end{tabular}
\end{table}

The current bound from the ATLAS result for the triple Higgs coupling scaling factor is given by $-0.4<\kappa_3<6.3$\cite{ATLAS:2022jtk}. The expected precision for the triple Higgs boson coupling measurements at the HL-LHC and at the three stages, 250GeV, 500GeV, and 1TeV, of the ILC are presented in table~\ref{tab-2}.

\begin{figure}[ht]
\centering
\includegraphics[width=\textwidth]{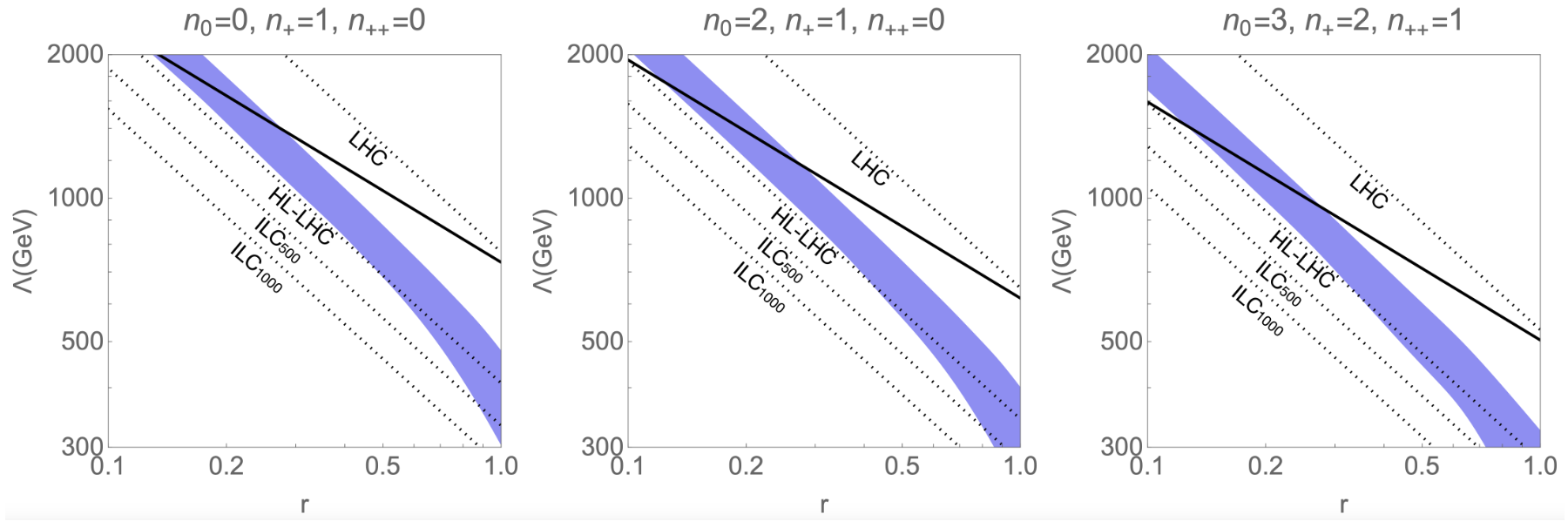}
\caption{Scenario where $\kappa_3$ is measured as the standard model value, $\kappa_3=1$, with the precisions shown in table~\ref{tab-2}. The dashed lines represent the respective upper bounds. The data is taken from \cite{DiMicco:2019ngk}.}
\label{fig-2}
\end{figure}

In figure~\ref{fig-2} the scenario where the triple Higgs coupling is measured to the SM value is considered for the various collider stages. The blue region represents the SFOPT allowed space, while the dashed lines represent the line for which the area below is allowed experimentally in this scenario. Since all the lines are on top or above the blue area, except for the last stage of 1TeV of the ILC, we conclude that, in this scenario, only by the last stage of the ILC the SFOPT could be ruled out.

\begin{figure}[ht]
\centering
\includegraphics[width=\textwidth]{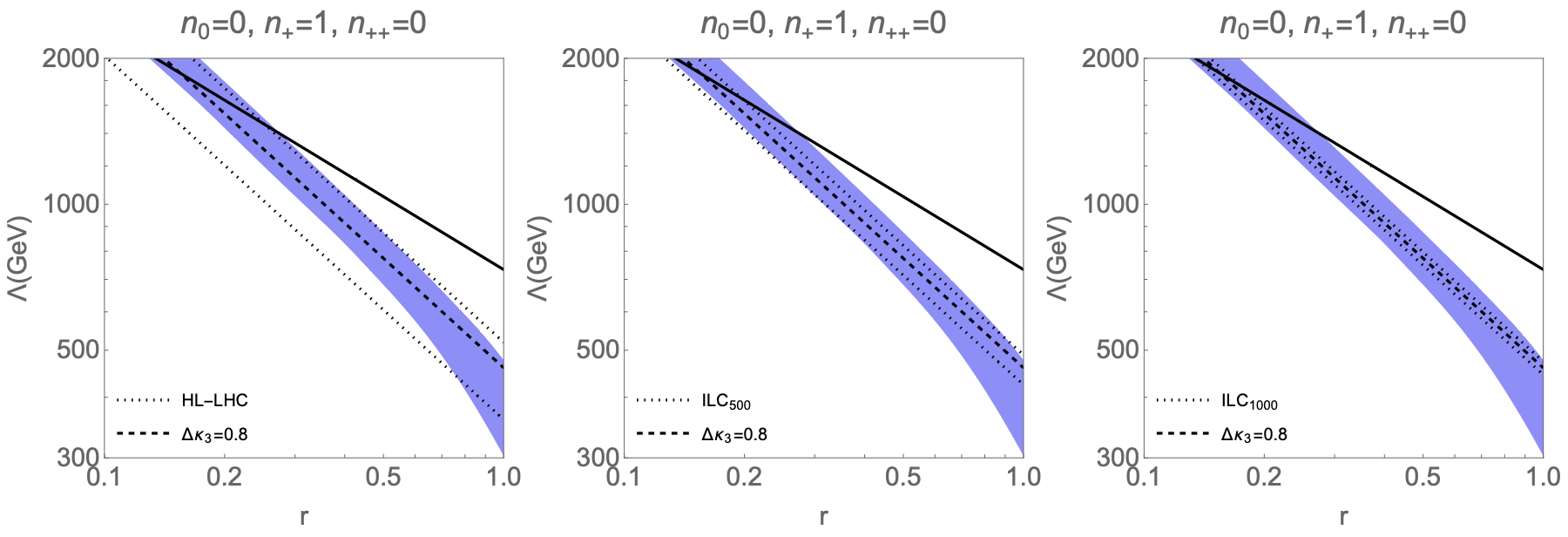}
\caption{Scenario where $\kappa_3$ is measured at the ideal value for a SFOPT, $\kappa_3=1.8$, with the precisions shown in table~\ref{tab-2}. The large dashed line represents the central value, while the smaller dashed lines represent the respective bounds.}
\label{fig-3}
\end{figure}

In figure~\ref{fig-3} the scenario where the triple Higgs coupling is measured at its ideal value for a SFOPT, $\kappa_3=1.8$, is considered for the various collider stages. The large dashed line represents the central value, while the smaller dashed lines bound the allowed area. Again, it can be seen that only the last stage of the ILC will be able to confirm the SFOPT by use of the triple Higgs coupling measurements alone. Note that this only happens for this ideal $\kappa_3$ value. Slight deviation would make it impossible to confirm the SFOPT even in the last ILC stage.

\begin{figure}[ht]
\centering
\includegraphics[width=0.8\textwidth]{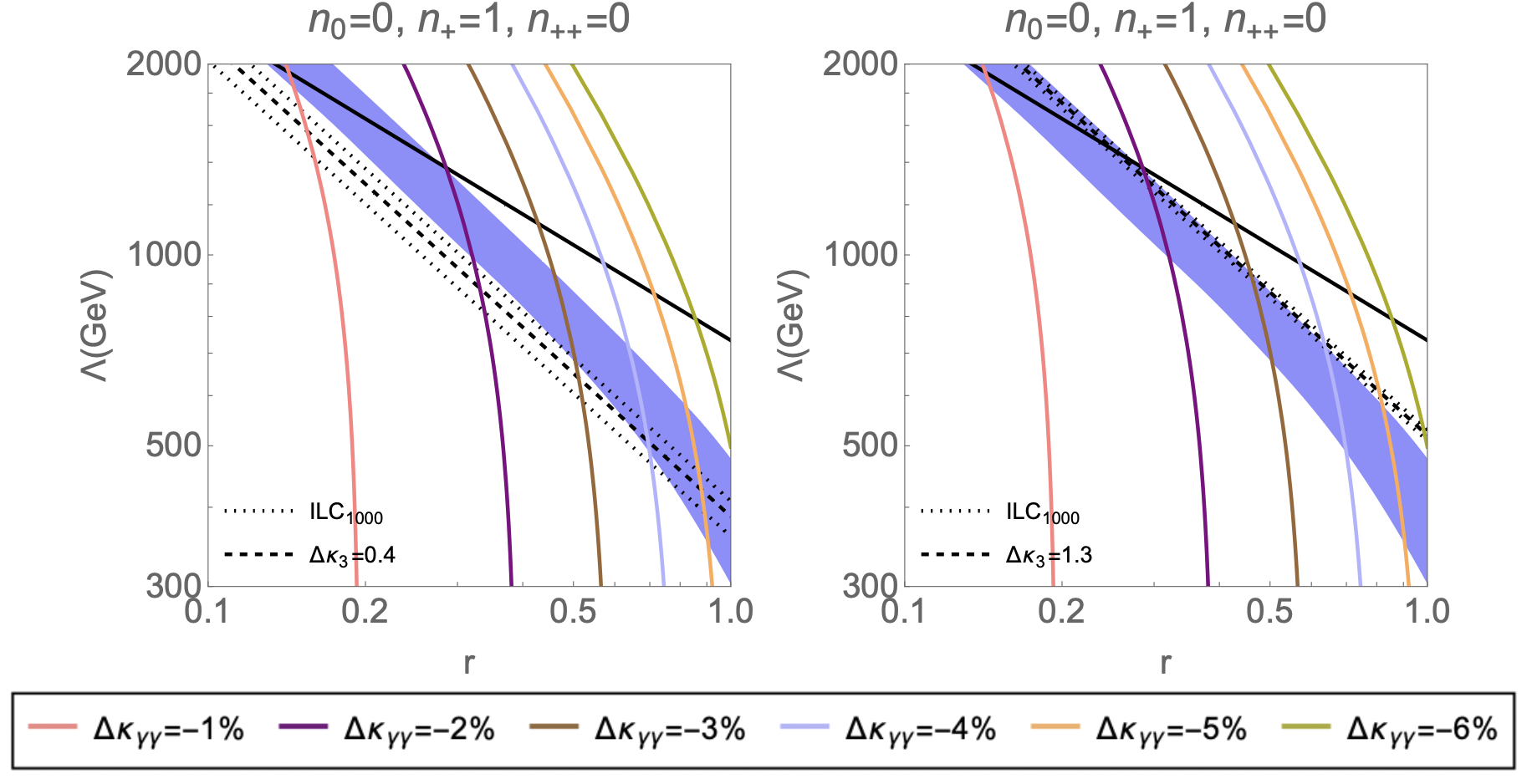}
\caption{Two example scenarios, $\kappa_3=1.4$(left), $\kappa_3=2.3$(right), where determination of the nature of the SFOPT is impossible by the triple Higgs coupling alone. The $\kappa_{\gamma\gamma}$ contours show how the $h\to\gamma\gamma$ measurements can remedy this problem.}
\label{fig-4}
\end{figure}

In figure~\ref{fig-4} two scenarios are considered, where the triple Higgs coupling scaling factor is measured as $\kappa_3=1.4$ and $\kappa_3=2.3$ in the left and right of the figure respectively. These represent different scenarios where not even the most accurate precision of the triple Higgs coupling would be able to determine the nature of the EWPT, for both situations present a central value contour that includes both section with a SFOPT and without. The $\kappa_{\gamma\gamma}$ contours are shown on top, and reveal that measurements of $h\to\gamma\gamma$ can restrict the parameter space to a smaller section of the parameter space, such that, in conjunction with the $\kappa_3$ measurements, it can confirm or deny the SFOPT. This exemplifies once again the importance of the $h\to\gamma\gamma$ measurements in helping to probe the nature of the EWPT, in conjunction with the triple Higgs coupling measurements.

\section{Conclusion}

In this work the parameter space generating a strongly first order EWPT is studied within the naHEFT framework. The sphaleron decoupling and completion conditions are imposed to said parameter space, and the parameter space generating detectable GW and PBH has been confirmed.

In section 4 this parameter space is studied for six different benchmark models. In figure \ref{fig-1} it can be seen that while PBH and GW at LISA are only able to probe a small section of the allowed parameter space, DECIGO can test a larger area. Contours on the triple Higgs coupling are show, and it can be seen that it strongly correlates with the existence of a SFOPT. Nonetheless, the $h\to\gamma\gamma$ is shown as a crucial complement to this study. In table \ref{tab-1} an example scenario of $h\to\gamma\gamma$ measurement is used in order to put constrains on the triple Higgs coupling values to generate a SFOPT. Both a bound required bound, outside of which the EWPT cannot be strongly first order, and a conservative bound, within which the EWPT is definitely strongly first order, are shown.

In section 5, the consequences of this study on the determination of the nature of the EWPT at future colliders is discussed. Namely, the HL-LHC and the three stages of the ILC are considered. It is shown that for very specific central values of the triple Higgs coupling, this measurement alone can determine the nature of the EWPT only at the last stage of the ILC. Outside this special cases, not even the most precise measurement of the triple Higgs coupling can conclusively determine the nature of the EWPT. Furthermore, it is discussed how the $h\to\gamma\gamma$ measurements can alleviate this issue. It is concluded that $h\to\gamma\gamma$ measurements will play a crucial role, alongside the triple Higgs coupling measurements, in testing the viability of various models to produce a SFOPT at the electroweak symmetry breaking in the early universe and, consequently, in successfully explaining BAU through EWBG.

\bibliography{FKT_references}

\end{document}